\documentclass[twocolumn]{aastex631}
\usepackage{amsmath}
\usepackage{hyperref}
\usepackage{mathrsfs}
\newcommand{\fermi}{\textit{Fermi}}
\newcommand{\psr}{SGR 0501$+$4516}
\newcommand{\lhaasosrc}{1LHAASO J0500$+$4454}
\newcommand{\fermisrc}{4FGL J0501.7$+$4459}

\shortauthors{Alford et al.}

\begin{document}

\title{Evidence of an Energetic Magnetar Powering 1LHAASO J0500$+$4454}

\correspondingauthor{Jason A. J. Alford}
\email{alford@nyu.edu}

\author[0000-0002-2312-8539]{J. A. J. Alford}
\affiliation{Division of Science, New York University Abu Dhabi, PO Box 129188, Abu Dhabi, UAE}
\affiliation{Center for Astrophysics and Space Science (CASS), New York University Abu Dhabi, PO Box 129188, Abu Dhabi, UAE}
\author[0000-0003-4679-1058]{J. D. Gelfand}
\affiliation{Division of Science, New York University Abu Dhabi, PO Box 129188, Abu Dhabi, UAE}
\affiliation{Center for Astrophysics and Space Science (CASS), New York University Abu Dhabi, PO Box 129188, Abu Dhabi, UAE}
\author[0000-0002-4441-7081]{M. Abdelmaguid}
\affiliation{Division of Science, New York University Abu Dhabi, PO Box 129188, Abu Dhabi, UAE}
\affiliation{Center for Astrophysics and Space Science (CASS), New York University Abu Dhabi, PO Box 129188, Abu Dhabi, UAE}
\author[0000-0002-6986-6756]{P. Slane}
\affiliation{Harvard-Smithsonian Center for Astrophysics, 60 Garden Street, Cambridge, MA 02138, USA}

\begin{abstract}  
We investigate the origin of the unidentified, extended TeV source 1LHAASO J0500$+$4454, considering three possible origins: cosmic rays interacting with a molecular cloud (MC), particles accelerated in a currently undetected supernova remnant (SNR), and an energetic outflow powered by a pulsar.
Upper limits on the CO and X-ray emission from the $\gamma$-ray emitting region disfavor the MC and SNR scenarios, respectively.
If a nebula of inverse Compton scattering $e^{\pm}$ powers 1LHAASO J0500$+$4454,
then spectral energy distribution modeling indicates that the current particle energy in the nebula is $\sim 4 \times  10^{48}$~erg.  
If the coincident magnetar \psr's rotational energy powered 1LHAASO J0500$+$4454, then a conservative energy budget calculation requires an initial magnetar spin period $P_{0} \lesssim 5$~ms and a spin-down timescale $\tau_{\rm sd} \lesssim 30$~yr, which has implications for the origins of magnetars.
\end{abstract}

\section{Introduction}
Surveys of the $\gamma$-ray sky are pushing the frontier to higher photon energies, and discovering lower surface brightness $\gamma$-ray sources. 
Recently, \cite{cao2024} published the 1LHAASO catalog of 90 very-high-energy (VHE, $E>0.1$~TeV) $\gamma$-ray sources. 
At least 35 1LHAASO sources are associated with pulsars and pulsar wind nebulae (PWNe).
\lhaasosrc \ is one of 15 unidentified TeV sources in the 1LHAASO catalog, and is located toward the relatively uncrowded Galactic anticenter region,  where it may be easier to identify its origin.
Within \lhaasosrc's $\theta_{\gamma} = 0.\!^{\circ}41 \pm 0.\!^{\circ}07$ radial extent lies the (also unidentified) \fermi \ source \fermisrc \ \citep{bal2023}.

Identifying the origins of $\gamma$-ray sources requires distinguishing between `leptonic' or `hadronic' $\gamma$-ray production, according to the classification of the particles producing the majority of the $\gamma$-rays.
In the hadronic scenario, high-energy protons collide with other protons in a dense medium (e.g. a molecular cloud, MC) to produce neutral pions ($\pi^{0}$), which then decay into the observed $\gamma$-rays.
High-energy $p^{+}$ are believed to be accelerated in SNRs, so MCs near SNRs are prime candidates for producing $\gamma$-ray sources \citep{ack2013,aha2013}.
In the leptonic scenario, high-energy e$^{\pm}$ produce $\gamma$-rays through inverse Compton scattering photons from the Cosmic Microwave Background (CMB) and other photon fields.
High-energy e$^{\pm}$ are produced in SNRs, pulsar magnetospheres, and at the termination shocks of PWNe, so both SNRs and pulsars are plausible sources of leptonic $\gamma$-rays.

\begin{figure*}
\centering
\includegraphics[width=1.0\linewidth]{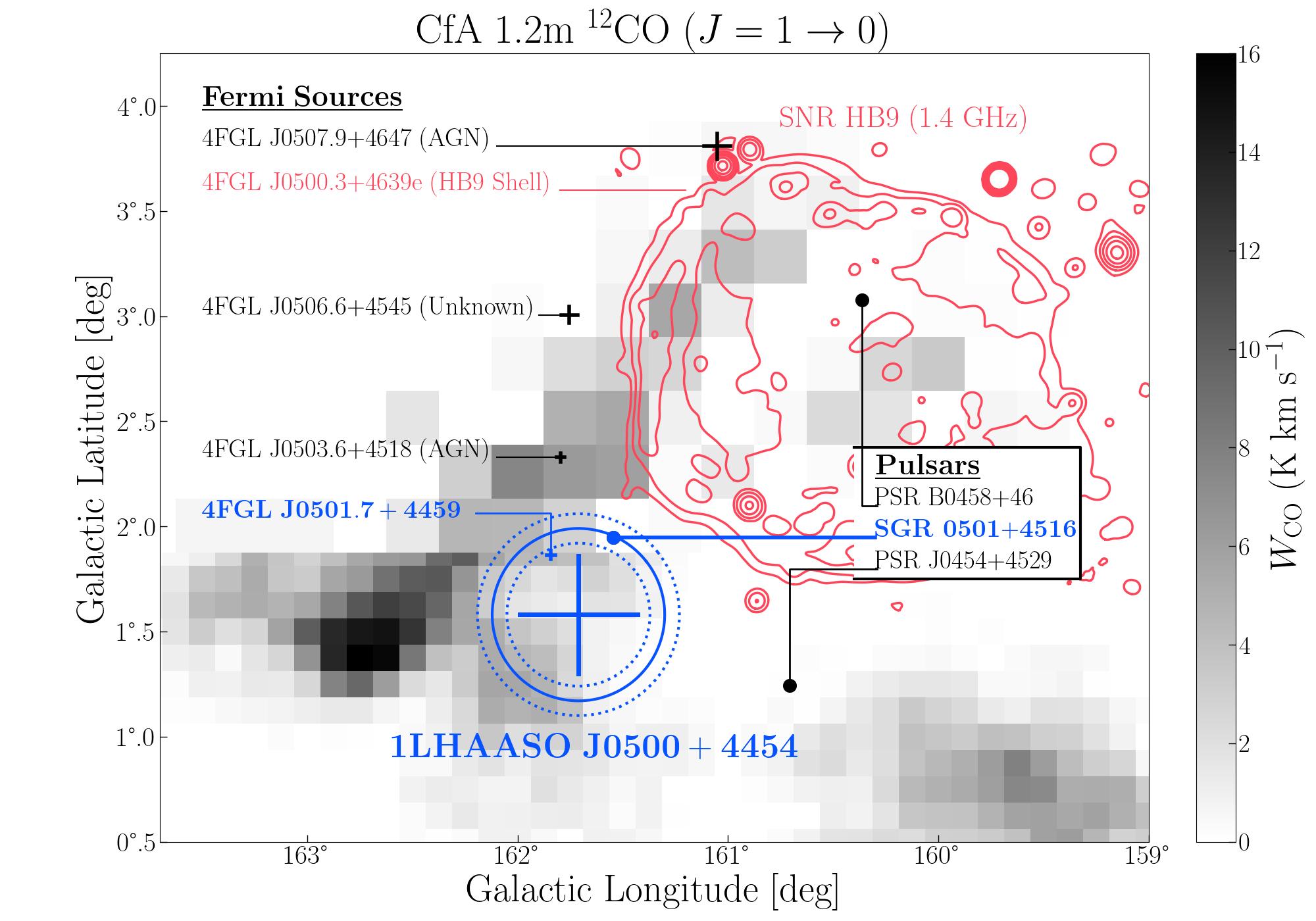}
\caption{$^{12}$CO ($J=1\rightarrow 0$) map of the region around \lhaasosrc, integrated in the velocity range $-7.15$ to $4.55$~km s$^{-1}$.
The solid blue circle indicates the angular extent of \lhaasosrc, $\theta_{\gamma} = 0.\!^{\circ}41 \pm 0.\!^{\circ}07$, and the dotted blue circles indicate the uncertainty in $R_{\gamma}$.
The blue crosshairs indicate the positional uncertainty of \lhaasosrc \ at the $95\%$ confidence level.
Black crosshairs indicate the positions (at $95\%$ confidence) of \fermi \ 4FGL-DR4 sources in the field, and associations listed in the \fermi \ 4FGL-DR4 catalog are indicated in parentheses.
All known pulsars and SNRs in the field are indicated.
Red contours trace the 1.4 GHz radio shell of the supernova remnant HB9.
}
\label{fig:region_map}
\end{figure*}

The magnetar \psr \ is the only known pulsar located within the extent of \lhaasosrc \ (see Figure \ref{fig:region_map}), raising the possibility of a physical association.
Unlike rotationally powered pulsars, magnetars are not often associated with extended VHE $\gamma$-ray sources, though some associations have been proposed \citep{hal2010,hess2018,got2019}, and young magnetars might have sufficient rotational and/or magnetic energy budgets \citep{dun1992,bel2016,kas2017}.
\psr's spin parameters and associated physical properties are listed in Table \ref{table:nearby_pulsars}.
No known SNRs or dense molecular clouds overlap \lhaasosrc.

This paper is organized as follows:
Section \ref{section:physical_properties} outlines the basic physical properties of \lhaasosrc \ and its environment.
Section \ref{section:SED} presents leptonic and hadronic models of the spectral energy distribution (SED) of \lhaasosrc \ and the associated Fermi source.
Section \ref{section:possible_power_sources} investigates possible origins of \lhaasosrc, and finds evidence for a PWN origin.  
Section \ref{section:energy_budget} discusses the PWN energy budget, if \lhaasosrc \ is a magnetar-powered PWN.
Section \ref{section:discusion} discusses our results in relation to previous work, and discusses the implications of magnetar TeV emission for magnetar birth properties and origins. 

\section{Physical Properties of the \lhaasosrc \ Region}
\label{section:physical_properties}

\begin{deluxetable*}{lcccccc}
\tablecaption{Known Pulsars Near \lhaasosrc}
\label{table:nearby_pulsars}
\tablewidth{0pt}
\tablehead{
\colhead{Pulsar} &  \colhead{$P$} & \colhead{$\dot{P}$} & \colhead{$\dot{E}$} & \colhead{$\tau_{\rm char}$} & \colhead{$B_s$} & References \\
 & (s)  & ($10^{-15}$~s~s$^{-1}$) & ($10^{31}$~erg~s$^{-1}$) & (kyr) & ($10^{12}$~G) & 
}
\startdata
\psr & $\approx5.76$ & $\approx5900$ & 120 & 15 & 190 & (1) \\ 
PSR J0454+4529 & $\approx1.39$ & $\approx4.9$  & $7.8$ & 4500 & 2.5& (2)\\
PSR B0458+46 & $\approx0.64$ & $\approx5.6$  & 84 & 1800 & 1.1 & (3) \\
\enddata
\tablecomments{References: (1) \cite{cam2014} (2) \cite{tan2020} (3) \cite{hob2004}}
\end{deluxetable*}

\begin{figure*}
\centering
\includegraphics[width=1.0\linewidth]{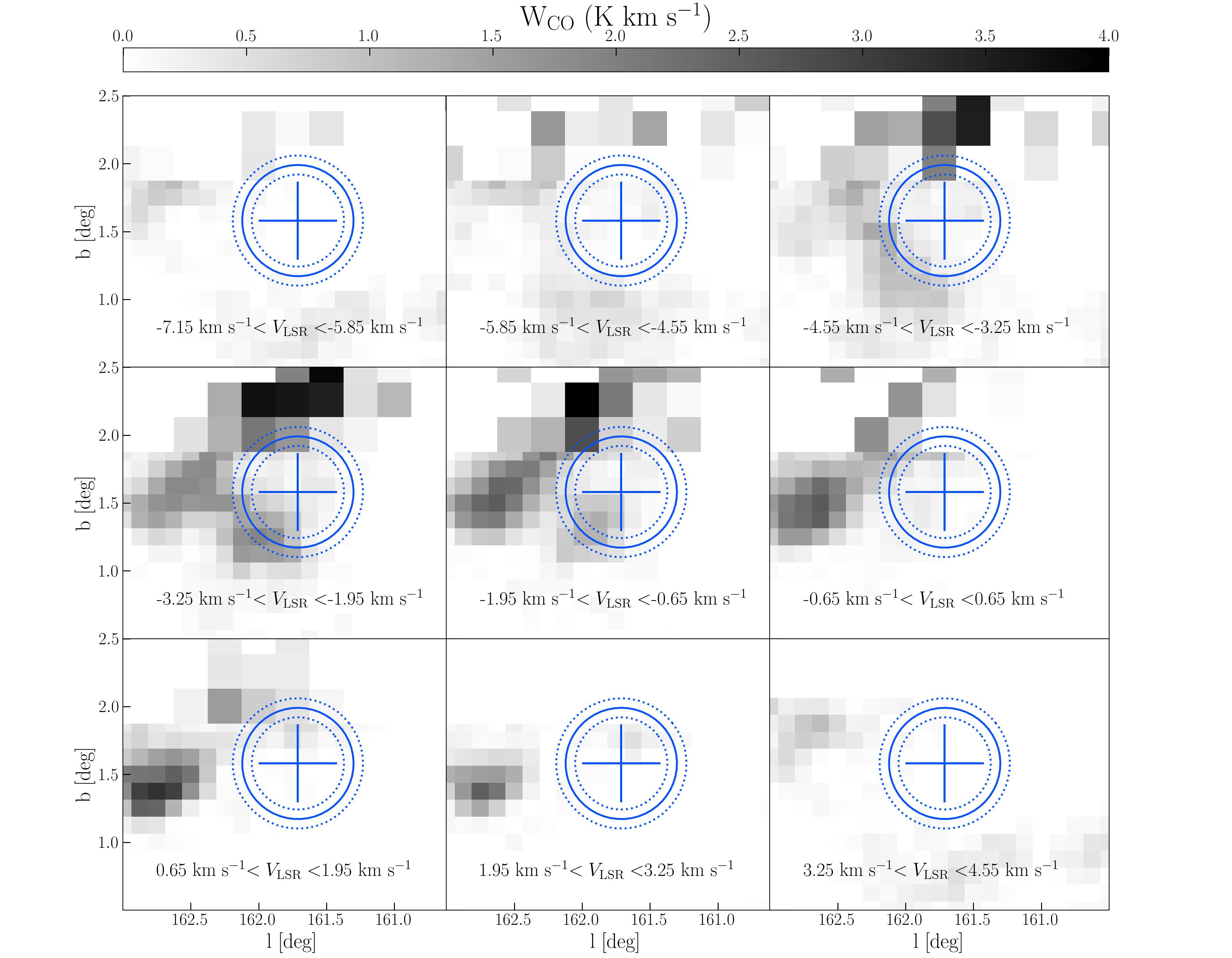}
\caption{$^{12}$CO ($J=1\rightarrow 0$) maps of the region around \lhaasosrc, integrated in various velocity ranges.
The solid blue circle indicates the angular extent of \lhaasosrc \  $\theta_{\gamma} = 0.\!^{\circ}41 \pm 0.\!^{\circ}07$, and the dotted blue circles indicate the uncertainty in $\theta_{\gamma}$.
The blue crosshairs indicate the positional uncertainty of \lhaasosrc \ at the $95\%$ confidence level.
}
\label{fig:cloud}
\end{figure*}

\subsection{Distance and Environment}
\lhaasosrc \ is located at R.A.$=75.\!^{\circ}01$, Dec.=$44.\!^{\circ}92$  ($l=161.\!^{\circ}71$, $b=1.\!^{\circ}58$), toward the Galactic anticenter region and  $1.\!^{\circ}6$ above the Galactic plane, implying a distance of $D_{\gamma}\sim2$~kpc if located in the Perseus Arm of the Milky Way \citep{xu2006}.
\lhaasosrc's radial extent $\theta_{\gamma} = 0.\!^{\circ}41 \pm 0.\!^{\circ}07$ corresponds to a physical radius 
\begin{equation}
R_{\gamma} = 14.3^{+2.4}_{-2.4}~\left(\frac{D_{\gamma}}{2~\rm{kpc}}\right)~\rm{pc}.
\label{eqn:lhaaso_radius}
\end{equation}
Figure \ref{fig:region_map} shows the position of \lhaasosrc \ in relation to all pulsars, SNRs, and Fermi sources listed in the ATNF Pulsar Catalog (version 2.1.1), Green's SNR catalog, and \fermi \ LAT 14 yr Source Catalog (4FGL-DR4), respectively \citep{man2005,gre2019,gre2025,bal2023}.
\psr \ is the only pulsar coincident with \lhaasosrc.
\cite{mon2018} measured an upper limit on \psr's proper motion that ruled out an association with the SNR HB9. 
Recently, \cite{chr25} measured  $\mu = 5.4~$mas~yr$^{-1}$, implying that \psr's birth site overlaps \lhaasosrc, if \psr's true age is comparable to its characteristic age ($\tau_{\rm char} = 15$~kyr, see also Figure \ref{fig:rosat}).

Figure \ref{fig:region_map} also shows the $^{12}$CO ($J=1\rightarrow0$) luminosity $L_{\rm co}$ in the region surrounding \lhaasosrc \ \citep{dam2022}.
Carbon monoxide (CO) emission traces molecular material \citep{tha1977}, and Figure \ref{fig:region_map} demonstrates that there is some molecular material coincident with \lhaasosrc. 
We calculate an upper limit on the molecular density within the \lhaasosrc \ region by assuming that all of the observed CO intensity $\mathrm{W}_{\rm co}$ is emitted from a hypothetical spherical cloud centered on the $\gamma$-ray source.
We integrate $W_{\rm co}$ within the $\theta_{\gamma} = 0^{\circ}.41$ extent of \lhaasosrc \ and the velocity range $-7.15~{\rm km~s^{-1}} < V_{\rm LSR} < 4.55~{\rm km~s^{-1}}$ (there is no significant CO emission within \lhaasosrc \ at other velocities covered in the CfA CO survey data), and find
\begin{equation}
\int \mathrm{W}_{\rm co} {\rm d}\Omega = 0.82~{\rm K~km~s^{-1}}~{\rm deg}^{2}.
\end{equation}
We convert the observed CO luminosity to an upper limit on a molecular cloud mass using a CO-to-H2 conversion factor of $2 \times 10^{20} \rm{cm^{-2} (K~km~s^{-1})^{-1}}$, which includes a factor of 1.36 to account for heavy elements \citep{bol2013,dam2022}:
\begin{equation}
M_{\rm cloud} \lesssim 4300~\left(\frac{\int W_{\rm co} {\rm d}\Omega}{\rm{0.82~K~km~s^{-1}~deg^{2}}}\right)  \left(\frac{D_{\gamma}}{2~\rm{kpc}}\right)^{2} \ M_{\odot}.
\end{equation}
The upper limit on proton number density $ n_{\rm cloud}$ of this hypothetical spherical cloud with mass $M_{\rm cloud}$ and radius $R_{\gamma}$ is:
\begin{eqnarray}
    n_{\rm cloud} \lesssim \left(\frac{M_{\rm cloud}}{\frac{4}{3} \pi R_{\gamma}^{3}}\right) m_{p^{+}}^{-1} \approx 14\pm7~\left(\frac{D_{\gamma}}{2~\rm{kpc}}\right)^{-1}~{\rm cm^{-3}}.
\label{eqn:co_density}
\end{eqnarray}

Figure \ref{fig:region_map} suggests that there is a shell of CO emission coincident with the boundary of \lhaasosrc \ and spanning velocities $-4.55~{\rm km~s^{-1}} < V_{\rm LSR} < 0.65~{\rm km~s^{-1}}$.
Figure \ref{fig:cloud} shows this same CO data in specific velocity ranges, so that the CO emission within \lhaasosrc \ is more easily seen (note the change in scale in Figure \ref{fig:region_map} versus Figure \ref{fig:cloud}).
The $\approx 14~{\rm cm^{-3}}$ cloud density estimated above is a conservative upper limit of the actual H density within \lhaasosrc, because it is the velocity-integrated value.

\subsection{Fermi LAT Analysis and $\gamma$-ray SED}
\label{section:physical_properties_sed}

\begin{figure*}
\centering
\includegraphics[width=1.0\linewidth]{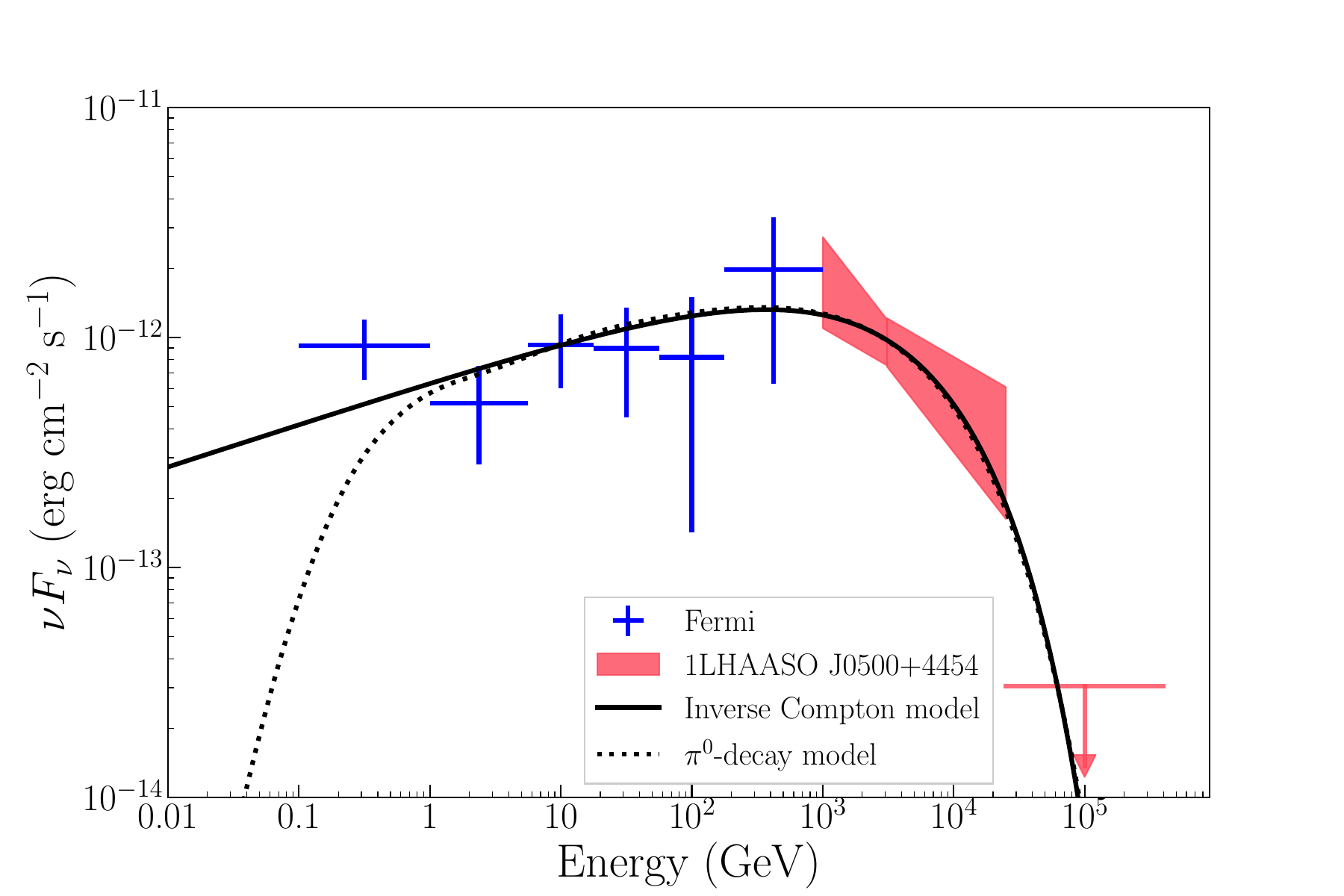}
\caption{Combined SED of \lhaasosrc \ and the associated \fermi \ source.
The solid curve shows the SED predicted by an inverse Compton (IC) model from an exponentially cutoff power-law electron population upscattering CMB photons.
The dashed curve shows the $\pi^{0}$-decay model. }
\label{fig:sed}
\end{figure*}

We selected photons within a 20 degree radius centered on \lhaasosrc \ (R.A.$=75.\!^{\circ}04$, Dec.$=44.\!^{\circ}91$), with photon energies ranging from 100~MeV to 500~GeV, and zenith angles less than 90$^{\circ}$.
The data were binned to $0.\!^{\circ}05$ per pixel.
We used the ${\tt gill\_iem\_v07}$ and ${\tt iso\_P8R3\_SOURCE\_V3\_v1}$ templates to model the Galactic diﬀuse and isotropic backgrounds, respectively. 
We first fit the \fermi \ LAT 14 yr Source Catalog (4FGL-DR4) to the data, and then re-fit after removing 4FGL-DR4 point source \fermisrc \ and replacing it with a source with the same position and extension as \lhaasosrc.
We find  a $3 \sigma$ preference for the later, extended morphology of \lhaasosrc. 
Figure \ref{fig:sed} shows the SED of this Fermi source.

Figure \ref{fig:sed} also shows the \lhaasosrc \ SED. 
We obtained the \lhaasosrc \ SED data from the First LHAASO Catalog of Gamma-Ray Sources, \cite{cao2024}. 
The \lhaasosrc \ SED in the $1$-$25$~TeV band is a power law $\frac{\mathrm{d}N}{\mathrm{d}E}  = N_0 \left(\frac{E}{3~{\rm TeV}}\right)^{-\Gamma}$ with $N_0 = 0.69\pm0.16 \times 10^{-13}~{\rm cm}^{-2}~{\rm s}^{-1}~\rm{TeV}^{-1}$ and $\Gamma=2.53\pm0.20$.
\lhaasosrc \ was not detected above $25$~TeV, with an upper limit $N_0 < 0.09 \times 10^{-16}~{\rm cm}^{-2}~{\rm s}^{-1}~\rm{TeV}^{-1}$ calculated for a source with the same radius, position, and power-law spectrum $\frac{\mathrm{d}N}{\mathrm{d}E} = N_{0} \left(\frac{E}{50~{\rm TeV}}\right)^{-2.53}$.

\subsection{Leptonic Cooling Timescales}
If the LHAASO TeV and Fermi GeV emission is leptonic, produced by $e^{\pm}$ inverse Compton scattering CMB photons, then the observed IC photon energies $E_{\gamma}$ correspond to $e^{\pm}$ energies $E_{e^\pm}$ through the relation \citep{ryb1979}:
\begin{equation}
E_{e^\pm} \approx 63~\sqrt{\frac{E_{\gamma}}{10 \ {\rm TeV}}}~{\rm TeV}.
\label{eq:ic_energy}
\end{equation}

Equation \ref{eq:ic_energy} indicates that the LHAASO observed photon energy range $1~{\rm TeV} \leq E_{\gamma} \leq 25$~TeV  corresponds to an e$^{\pm}$ energy range $20~{\rm TeV} \lesssim E_{e^\pm} \lesssim 100$~TeV.
The $e^{\pm}$ will lose energy through both inverse Compton scattering and synchrotron radiation.
Assuming the CMB dominates the photon energy density, an $e^{-}$ of energy 40 TeV upscattering CMB photons will lose half its energy on an inverse Compton cooling timescale
\begin{equation}
\label{eqn:ic_timescale}
\tau_{\rm IC}  \approx 30  \left(\frac{E_{e^\pm}}{40~{\rm TeV}}\right)^{-1}~{\rm kyr},
\end{equation}
and on a synchrotron cooling timescale
\begin{equation}
\label{eqn:sync_timescale}
\tau_{\rm sync} \approx   50 \left(\frac{E_{e^\pm}}{40~{\rm TeV}}\right)^{-1}  \left(\frac{B_{\perp}}{2 \mu{\rm G}} \right)^{-2}~{\rm kyr}.
\end{equation}
$\tau_{\rm sync}$ is less certain than $\tau_{\rm IC}$ because of its $\tau_{\rm sync} \propto B_{\perp}^{-2}$ dependence.
We therefore adopt $\tau_{\rm IC} \approx 30$~kyr as a conservative upper limit on the age of $\gamma-$ray emitting e$^{\pm}$ in \lhaasosrc.

\section{SED Modeling}
\label{section:SED}
We modeled the SED of \lhaasosrc \ (presented in Section \ref{section:physical_properties_sed}) in order to calculate $W$, the total particle energy in this TeV source.
We performed this SED modeling using {\tt Naima} \citep{zab2015}. 
Section \ref{subsection:SED_leptonic} discusses the leptonic model, and Section \ref{subsection:SED_hadronic} discusses the hadronic model.

\begin{deluxetable*}{lccccc}
\tablecaption{SED Modeling Results}
\label{table:SED}
\tablewidth{0pt}
\tablehead{
\colhead{Model} &  \colhead{$\frac{\mathrm{d}N}{\mathrm{d}E}$} & \colhead{$A~[\mathrm{eV}^{-1}]$} & \colhead{$\alpha$} & \colhead{$E_{\rm cut}~[\mathrm{TeV}]$} & \colhead{$\chi^{2}_{\nu}$ (d.o.f.)} 
}
\startdata
Leptonic (Inverse Compton) & $A \left(\frac{E_{e^{\pm}}}{1~\mathrm{TeV}}\right)^{-\alpha} \exp{\left(\frac{-E_{e^{\pm}}}{{E_{\rm cut}}}\right)}$ &  $(1.95 \pm 0.27) \times 10^{34}$ & $2.63 \pm 0.08$ & $37 \pm 7$ & 0.6 (9) \\
\hline
Hadronic ($\pi^{0}$-decay) & $A \left(\frac{E_{p^{+}}}{1~\mathrm{TeV}}\right)^{-\alpha}   \exp{\left(\frac{-E_{p^{+}}}{{E_{\rm cut}}}\right)}$ & $(8.9 \pm 1.4) \times 10^{34}$ & $1.91 \pm 0.07$ & $70 \pm 19$ & 0.9 (9) \\
\enddata
\tablecomments{Results of fitting leptonic and hadronic SED models to the $\gamma-$ray emission from the \lhaasosrc region, including data from Fermi LAT, LHAASO WCDA, and LHAASO KM2A detectors (Figure \ref{fig:sed}). 
The leptonic and hadronic models are powered by exponentially cutoff power-law particle distributions $\frac{\mathrm{d}N}{\mathrm{d}E}$. 
$1\sigma$ errors on the model parameters are indicated.}
\end{deluxetable*}

\subsection{Leptonic (IC) Model}
\label{subsection:SED_leptonic}

We fit the combined Fermi LAT and \lhaasosrc \ SED with an inverse Compton emission model where relativistic $e^{\pm}$ upscatter CMB seed photons.
We explored both pure power law and exponentially cutoff power law particle spectra, and found that an exponentially cutoff power law 
\begin{equation}
\frac{\mathrm{d}N}{\mathrm{d}E} = A \left(\frac{E_{e^{\pm}}}{1~\mathrm{TeV}}\right)^{-\alpha} \exp{\left(\frac{-E_{e^{\pm}}}{{E_{\rm cut}}}\right)} \ {\rm eV}^{-1}
\end{equation}
 is required to fit the data.
The best fit parameters are listed in Table \ref{table:SED}, with $1\sigma$ errors computed from the covariance matrix.
Figure \ref{fig:sed} shows the SED predicted by this model and the observed SED.
We calculate the total $e^{\pm}$ energy $W_{e^{\pm}}$  ($\equiv \int_{E_{\rm min}}^{E_{\rm max}} E_{e^{\pm}} \frac{\mathrm{d}N}{\mathrm{d}E} {\rm dE_{e^{\pm}}}$)
by setting $E_{\rm max}=510$~TeV (the {\tt Naima} default value and $\gg E_{\rm cut}$) and leaving $E_{\rm min}$ as a free parameter:
\begin{equation}
\label{eq:W_e}
W_{e^{\pm}} \approx 3.8^{+2.7}_{-1.6} \times 10^{48}~\left(\frac{E_{\rm min}}{1~\mathrm{GeV}}\right)^{2 - \alpha}~\left(\frac{D_{\gamma}}{2~\rm{kpc}}\right)^2~\rm{erg}.
\end{equation}
The $1\sigma$ errors in $W_{e^{\pm}}$ (and in $W_{p^{+}}$ below) were computed by Monte Carlo sampling. 
The total number of particles $N$ ($\equiv \int_{E_{\rm min}}^{E_{\rm max}}  \frac{\mathrm{d}N}{\mathrm{d}E} {\rm dE_{e^{\pm}}}$)  in this energy range is:
\begin{equation}
\label{eq:N}
N_{\pm} \approx 9_{-3}^{+6} \times 10^{50} \left(\frac{E_{\rm min}}{1~\mathrm{GeV}}\right)^{1-\alpha}~\left(\frac{D_{\gamma}}{2~\rm{kpc}}\right)^2.
\end{equation}

\subsection{Hadronic ($\pi^{0}$-decay) Model}
\label{subsection:SED_hadronic}
We consider a spectrum $\frac{\mathrm{d}N}{\mathrm{d}E}$ of high-energy $p^{+}$ producing $\pi^0$ through collisions with low-energy $p^{+}$ in a molecular cloud of density $n_{\rm cloud}$.
We model the high-energy $p^{+}$ particle spectrum $\frac{\mathrm{d}N} {\mathrm{d}E}$ with an exponentially cutoff power law:
\begin{equation}
\frac{\mathrm{d}N}{\mathrm{d}E} = A \left(\frac{E_{p^{+}}}{1~\mathrm{TeV}}\right)^{-\alpha}   \exp{\left(\frac{-E_{p^{+}}}{{E_{\rm cut}}}\right)} \ {\rm eV}^{-1}.
\end{equation}
In this model $A \propto n_{\rm cloud}^{-1} D_{\gamma}^2$, which we fix at fiducial values $n_{\rm cloud} = 14~{\rm cm}^{-3}$ (see Equation \ref{eqn:co_density}) and $D_{\gamma} = 2$~kpc.
The best fit parameters are listed in Table \ref{table:SED}.
We calculate the total $p^{+}$ energy $W_{p^{+}}$ ($\equiv \int_{E_{\rm min}}^{E_{\rm max}} E_{p^{+}} \frac{\mathrm{d}N}{\mathrm{d}E} {\rm dE_{p^{+}}}$) by setting 
$E_{\rm min} = 1.22$~GeV, the dynamical threshold for $\pi^0$ production, and we set $E_{\rm max} = 10$~PeV (the {\tt Naima} default value and $\gg E_{\rm cut}$).
We find: 
\begin{equation}
\label{eq:W_p}
W_{p^{+}} = 1.4^{+0.2}_{-0.2}  \times 10^{48} ~ \left(\frac{n_{\rm cloud}}{14~{\rm cm}^{-3}}\right)^{-1}~\left(\frac{D_{\gamma}}{2~\rm{kpc}}\right)^2~\rm{erg}.
\end{equation}
The $1\sigma$ errors in $W_{p^{+}}$ were computed by Monte Carlo sampling.

\begin{figure*}
\centering
\includegraphics[width=0.9\linewidth]{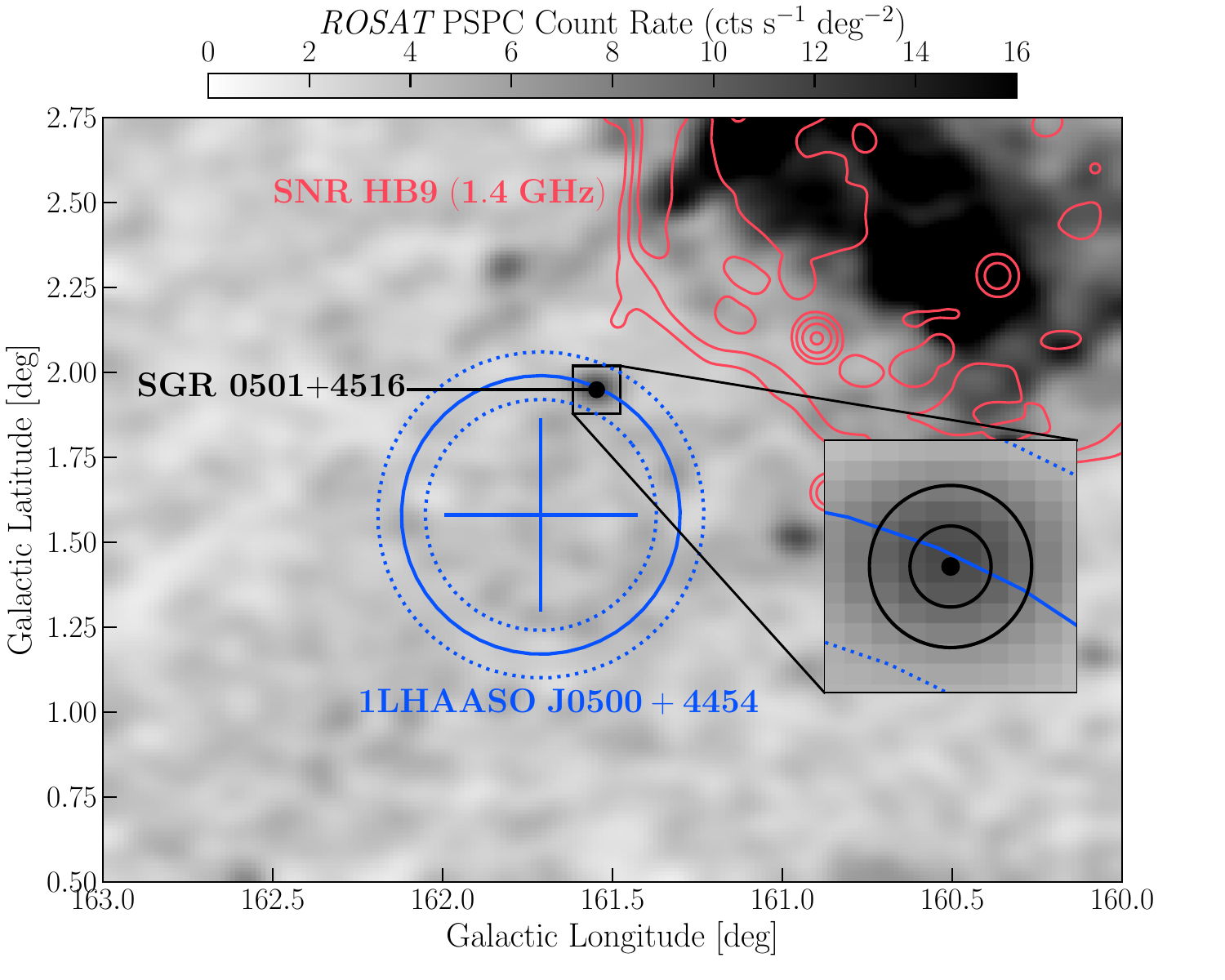}
\caption{ROSAT PSPC observations of \lhaasosrc \ and the surrounding region.
Red contours indicate the radio emission from SNR HB9.
We have indicated the position of \psr, which is clearly detected by ROSAT at the edge of \lhaasosrc.
Inset: Circles around \psr \ indicate its possible birth locations for pulsar ages $\tau_{psr} = \tau_{\rm char}$ and $\tau_{psr} = 2 \tau_{\rm char}$ ($\tau_{\rm char}=15$~kyr), and proper motion $\mu = 5.4~$mas~yr$^{-1}$.}
\label{fig:rosat}
\end{figure*}

\section{What powers \lhaasosrc?}
\label{section:possible_power_sources}
\lhaasosrc's large radial extension $\theta_{\gamma} = 0.\!^{\circ}41 \pm 0.\!^{\circ}07$ rules out an extra-Galactic origin.
Here we discuss three plausible Galactic origins for \lhaasosrc: a molecular cloud  (Section \ref{section:possible_power_sources:MC}), a supernova remnant  (Section \ref{section:possible_power_sources:SNR}), and an energetic outflow from a pulsar  (Section \ref{section:possible_power_sources:PSR}).
The plausibility of these scenarios depends on the energetics derived from SED modeling in Section \ref{section:SED} and also the locations of nearby MCs, SNRs, and pulsars (see Figure \ref{fig:region_map}).

\subsection{Molecular Cloud Scenario}
\label{section:possible_power_sources:MC}
If \lhaasosrc \ is powered by cosmic rays interacting with an MC, then there should be a sufficiently  dense MC coincident with the observed $\gamma$-rays. 
The nearest known SNR to \lhaasosrc \ is HB9, and is therefore the most likely source of such cosmic rays.
We will use $W_{p^{+}}$ (Equation \ref{eq:W_p}) to calculate the energy budget of this hadronic scenario, assuming cosmic ray $p^{+}$ accelerated by SNR HB9 are powering \lhaasosrc.
This cosmic ray proton energy currently with \lhaasosrc \ ($W_{p^{+}}$) and the projected distance from the center of \lhaasosrc \ to the center of HB9 imply a rate of energy injected from HB9 into cosmic ray $p^{+}$.
The calculation of this energy injection rate indicates that the low upper limit on the molecular density in the \lhaasosrc \ region, combined with its projected distance from HB9, implies an unrealistic cosmic ray energy budget.
This disfavors this hadronic scenario, and the details of these calculations are presented in Appendix \ref{appendix_mc}.

\subsection{SNR Scenario}

\label{section:possible_power_sources:SNR}
It is possible that \lhaasosrc \ is powered by particles accelerated to high energies in an SNR.
Figure \ref{fig:region_map} shows that the only known nearby SNR HB9 does not overlap the \lhaasosrc \ emission region.
However, a previously unknown SNR at this location could power \lhaasosrc.
The hypothetical SNR considered in this section is not assumed to be associated with any known astronomical objects in this field.

We list here the observed properties of \lhaasosrc \ that we will use to assess the plausibility of an SNR origin.
First, we assume that the \lhaasosrc \ radius  is greater than or equal to the SNR radius:
\begin{equation}
R_\gamma \gtrsim R_{\rm snr} 
\end{equation}
Second, Figure \ref{fig:rosat} shows archival ROSAT All Sky Survey observations (ROR numbers 190568 and 190906) of \lhaasosrc \ and the surrounding region .
Setting $n_{\rm H} \lesssim 5 \times 10^{21}~\rm{cm}^{-2}$, a typical upper limit on the Galactic column density in this direction \citep{HI2016}, the ROSAT $3 \sigma$ upper limit on the unabsorbed 0.1$-$2.5~keV flux from the region with the same position and extension as \lhaasosrc \ is: 
\begin{equation}
F_x \lesssim 1.0 \times 10^{-11}~{\rm erg~cm^{-2}~s^{-1}}.
\label{eq:rosat_flux}
\end{equation}

Third, we assume that the SNR has accelerated $e^{-}$ to $\sim40$~TeV and accelerated $p^{+}$ to $\sim70$~TeV.

These constraints on the size, X-ray flux, and particle acceleration within the \lhaasosrc \ region pose significant challenges for the SNR origin scenarios, for an SNR at any evolutionary stage.
The calculations demonstrating these challenges are presented in Appendix \ref{appendix_snr}.

\subsection{Pulsar Scenario}
\label{section:possible_power_sources:PSR}

A pulsar with an initial spin period $P_{0}$ and moment of inertia $I_{45} = 10^{45}$~g~cm$^{2}$ has an initial rotational energy
\begin{equation}
\label{eqn:e_rotation}
E_{\rm rot} = 1.97 \times 10^{50} \ I_{45} \ \left(\frac{P_0}{10~{\rm ms}}\right)^{-2}   \ {\rm erg},
\end{equation}
and an initial magnetic energy 
\begin{equation}
\label{eqn:e_mag}
E_{\rm B} = 1.66 \times  \left(\frac{B_{0}}{10^{16}~{\rm G}}\right)^{2}  \times 10^{49}{\rm erg}. 
\end{equation}
Equations \ref{eqn:e_rotation} and \ref{eqn:e_mag} indicate that a sufficiently rapidly rotating and/or magnetized pulsar has an energy budget greater than the current \lhaasosrc \ leptonic energy $W_{e^{\pm}} \approx 3.8^{+2.7}_{-1.6} \times 10^{48}$~erg (Equation \ref{eq:W_e}).
If $W_{e^{\pm}}$ originates from the initial rotational kinetic energy of the pulsar $E_{\rm rot}$ (Equation \ref{eqn:e_rotation}),
then $E_{\rm rot} > W_{e^{\pm}}$ requires an initial spin period $P_{0} \lesssim 70$~ms.
Likewise, if $W_{e^{\pm}}$ originates from the pulsar's initial magnetic energy, then $E_{\rm B} > W_{e^{\pm}}$ and Equation \ref{eqn:e_mag} requires internal magnetic field strength $B \gtrsim 5 \times 10^{15}$~G.

There are three known pulsars in the field of \lhaasosrc \ (Figure \ref{fig:region_map}):  PSR B0458$+$46, PSR J0454$+$4529, and the magnetar \psr.
PSR B0458$+$46's distance was recently measured by FAST to be $>2.7$~kpc, beyond the Perseus arm \citep{jin2023}.
At this distance PSR B0458$+$46's $2.\!^{\circ}0$ angular separation from  \lhaasosrc \ and the $30$~kyr upper limit on the age of \lhaasosrc \ corresponds to a tangential velocity $v_{\perp} \gtrsim 3700$~km~s$^{-1}$.
Such a large velocity is highly unlikely, given the measured proper motions of young pulsars \citep{hob2005}, and argues against an association with \lhaasosrc.

PSR J0454$+$4529 has spin parameters $P \approx 1.39$~s and $\dot{P} \approx 4.89 \times 10^{-15}$~s~s$^{-1}$ \citep{tan2020}. 
If its true age $\tau_{\rm true}$ is comparable to its characteristic age  $\tau_{char} \equiv P / (2 \dot{P}) \approx 4.5$~Myr, then it is too old to have powered \lhaasosrc, whose age should be $\lesssim \tau_{\rm IC}\approx30$~kyr in this scenario.
If PSR J0454$+$4529's true age $\tau_{\rm true}$ is actually $\lesssim 30$~kyr, then its initial spin period is comparable to its current spin period: $P_0 \approx P \approx 1.39$~s.
Then its total spin-down energy since birth $E_{\rm sd} \approx \tau_{\rm true} \dot{E}_0 \lesssim (30~{\rm kyr}) (8\times 10^{31}~{\rm erg~s^{-1}}) \approx 7.5 \times 10^{43}~{\rm erg}$, which is $<< W_{e^{\pm}} \sim 4 \times  10^{48}$~erg (Equation \ref{eq:W_e}) and insufficient to power \lhaasosrc.

\cite{mon2018} measured an upper limit on \psr's proper motion $\mu < 320~{\rm mas~yr^{-1}}$ ($90\%$ confidence level), which ruled out a proposed association with SNR HB9 (red contours in Figure \ref{fig:region_map}).
Recently, \cite{chr25} reported an Hubble Space Telescope-measured proper motion $\mu = 5.4 \pm 0.6~{\rm mas~yr^{-1}}$, consistent with \psr \ being born within \lhaasosrc's  $95\%$ positional uncertainty region (blue crosshairs in Figure \ref{fig:region_map}). 
Black circles centered on \psr \ in Figure \ref{fig:rosat} indicate radii of $15$~kyr $\times 5.4 \ {\rm mas~yr^{-1}}$ and $30$~kyr $\times 5.4 \ {\rm mas~yr^{-1}}$.
\psr's offset from the center of \lhaasosrc \ is comparable to the offsets of other middle-aged ($\sim 10$~kyr) pulsars powering TeV pulsar wind nebulae \citep{hess2018}.

Finally, \psr's 15~kyr characteristic age is consistent with the $\tau_{\rm IC}\approx30$~kyr upper limit (see Equation \ref{eqn:ic_timescale}) on the age of the particles in \lhaasosrc.
Since both the position and age of \psr \ are consistent with a source that could be powering \lhaasosrc, we next consider the required energy budget.

\section{Energy Budget of \psr}
\label{section:energy_budget}

\begin{figure*}
\centering
\includegraphics[width=1.0\linewidth]{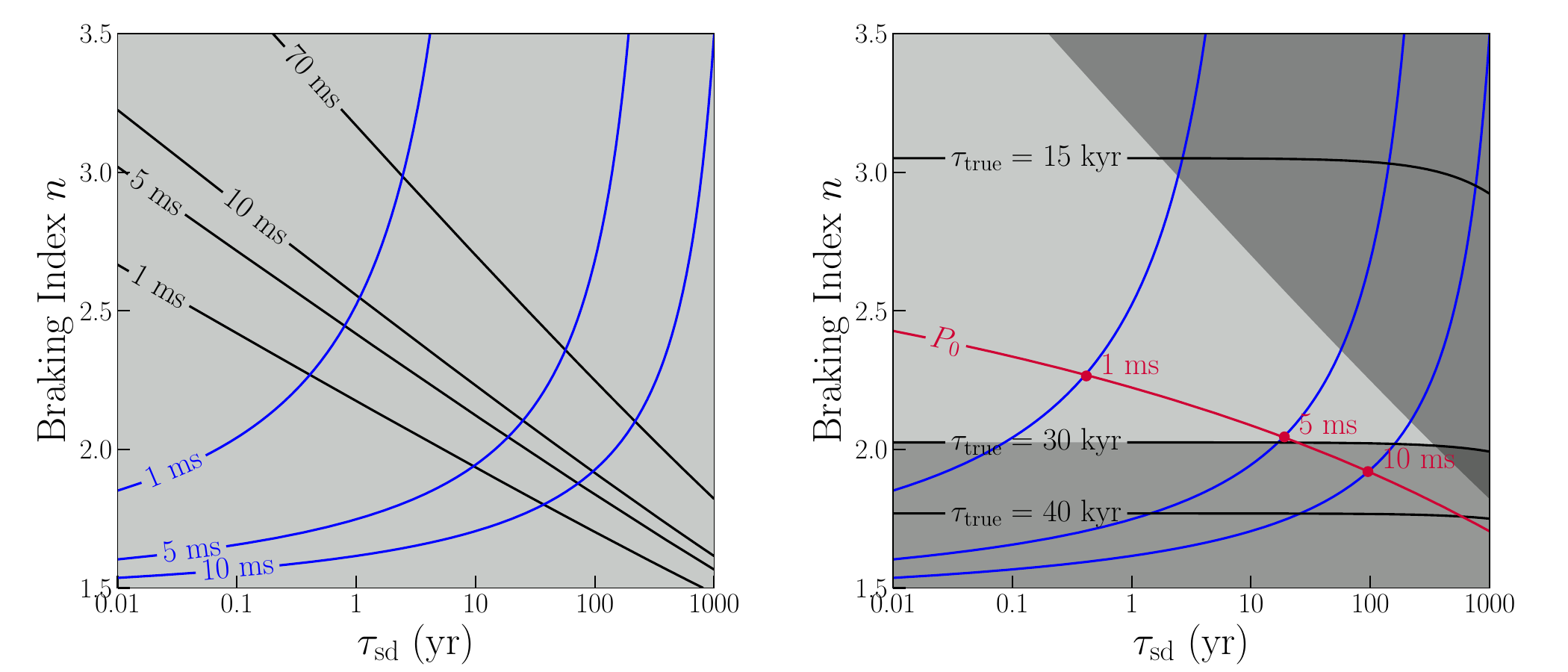}
\caption{{\it Left}: \psr's initial spin period $P_{0}$ as a function of  spin-down timescale $\tau_{\rm sd}$ and braking index $n$, calculated using two independent methods.
The black lines indicate curves of constant $P_0$ calculated with standard spin-down formalism (Equation \ref{eqn:period_0}), with \psr's currently measured $P=5.76$~s and $\dot{P} = 5.9 \times 10^{-12}$~s~s$^{-1}$.
The blue lines trace curves of constant $P_0$ calculated using the currently measured particle energy in \lhaasosrc \ (Equation \ref{eq:W_e}) and accounting for adiabatic energy losses.
{\it Right}: 
Lines of constant pulsar true age $\tau_{true}$ are indicated.  
$\tau_{true} > 30$~kyr is ruled out by the half life of $\approx40$~TeV $e^{\pm}$ inverse Compton scattering CMB photons (Equation \ref{eqn:ic_timescale}).
The red curve indicates where the two $P_{0}$ calculations in the left panel are consistent.
}
\label{fig:n_vs_tau}
\end{figure*}

Here we calculate constraints on \psr's initial spin period $P_{0}$, assuming that its initial rotational energy dominates the input energy budget for \lhaasosrc. 
\lhaasosrc's total $e^{\pm}$ energy $W_{e^{\pm}} \approx 3.8^{+2.7}_{-1.6} \times 10^{48}$~erg (Equation \ref{eq:W_e}).
The black curves in the left panel of Figure \ref{fig:n_vs_tau} show $P_{0}$ calculated for various values of the pulsar braking index $n$ and spin-down timescale $\tau_{\rm sd}$, calculated with the standard spin-down formalism:
\begin{eqnarray}
\label{eqn:period_0}
P & = 5.76~{\rm s} & = P_{0} \left( 1 + \frac{t}{\tau_{\rm sd}}     \right)^{\frac{1}{n-1}},
\end{eqnarray}
where we have set the time  $t$ equal to the pulsar's true age $\tau_{\rm true}$:
\begin{eqnarray}
\tau_{\rm true} = \frac{2 \tau_{\rm char}}{(n -1)} - \tau_{\rm sd}.
\end{eqnarray}
For the canonical dipole braking index value $n=3$, the requirement that $P_{0} \lesssim 70$~ms  corresponds to $\tau_{\rm sd} \lesssim 10$~yr, with larger values of $\tau_{\rm sd}$ allowed for $n<3$.

The above $P_{0}$ upper limit, calculated by setting $E_{\rm rot} > W_{e^{\pm}}$ is a hard upper bound on $P_{0}$, but energy losses would surely be significant over the lifetime of a pulsar wind nebula.
A more realistic $P_{0}$ upper limit would account for these energy losses and require that $E_{\rm rot} > E_{\rm PWN}$, where  $E_{\rm PWN}$ is the total energy injected into the $E_{\rm PWN}$ over its lifetime.
We will calculate this upper bound on $P_{0}$ by setting $E_{\rm rot} > E_{\rm PWN}(t=\tau_{\rm sd})$, where $E_{\rm PWN}(t=\tau_{\rm sd})$ is the total PWN particle energy at time $t=\tau_{\rm sd}$.
This is a conservative lower bound on $E_{\rm PWN}$ (upper bound on $P_{0}$), since according to standard pulsar spin-down formalism most of the pulsar's rotational energy is injected between time $t=0$ and time $t = \tau_{\rm sd}$: 
\begin{eqnarray}
\label{eqn:Edot_0}
\dot{E}(t) & = & \dot{E_{0}} \left( 1 + \frac{t}{\tau_{\rm sd}} \right)^{-\frac{(n+1)}{(n-1)}},
\end{eqnarray}
and we neglect energy injected after $t=\tau_{\rm sd}$.

We calculate $E_{\rm PWN}(t=\tau_{\rm sd})$ by accounting for adiabatic losses, which are expected to dominate the PWN energy budget at times $ t > \tau_{\rm sd}$ \citep{gel2009}.
The energy of a relativistic ideal gas (adiabatic index $\gamma = 4/3$) in a spherical PWN bubble is inversely proportional to the PWN radius: $E_{\rm PWN} \propto R_{\rm PWN}^{-1}$.  
The observed size of \lhaasosrc, gives the current PWN radius: $R_{\rm PWN}(t=\tau_{\rm true}) \approx 14.3$~pc.
$R_{\rm PWN}(t=\tau_{\rm sd})$ can be calculated using the well-known self-similar PWN expansion law \citep{che1977, rey1984}:

\begin{eqnarray}
R_{\rm PWN} = 1.1~\left(\frac{\dot{E_0}}{10^{38}~\rm{erg~s^{-1}} }\right)^{1/5}~\left(\frac{t}{1~\rm{kyr}} \right)^{6/5}~\rm{pc}.
\label{eqn:r_pwn}
\end{eqnarray}

$R_{\rm PWN}$ depends weakly on the SNR explosion energy and ejecta mass ($R_{\rm PWN} \propto E_{\rm SN}^{3/10} M_{\rm ej}^{-1/2}$), which we have set to  $E_{\rm SN} = 10^{51}~\rm{erg}$ and $M_{\rm ej} = 10~M_{\odot}$ in Equation \ref{eqn:r_pwn} above.
We calculate $R_{\rm PWN}$ for the range of $n$ and $\tau_{\rm sd}$ shown in the left panel of Figure  \ref{fig:n_vs_tau}. 
For each $n$ and $\tau_{\rm sd}$ we calculate an expansion ratio 
\begin{equation}
R_{\rm exp} \equiv \frac{R_{\rm PWN}(\tau_{\rm true})}{R_{\rm PWN}(\tau_{\rm sd})}
\end{equation}
which we use to calculate
\begin{equation}
E_{\rm PWN}(\tau_{\rm sd}) = R_{\rm exp} \times E_{\rm PWN}(\tau_{\rm true}),
\end{equation}
where $E_{\rm PWN}(\tau_{\rm true}) = W_{e^{\pm}}$ (Equation \ref{eq:W_e}).

For each $E_{\rm PWN}(\tau_{\rm sd})$, Equation \ref{eqn:e_rotation} gives the corresponding $P_{0}$.
Curves of constant $P_{0}$ calculated in this way are shown in blue in the left panel of Figure \ref{fig:n_vs_tau}. 
The red curve in the right panel of Figure \ref{fig:n_vs_tau} indicates the $n$ and $\tau_{\rm sd}$ values where $P_{0}$ calculated in the spin-down formalism (black in the left panel) and $P_{0}$ calculated from adiabatic losses (blue in the left panel) are consistent.
The dark shaded regions in the right panel of Figure  \ref{fig:n_vs_tau} indicate parameters that are ruled out by the requirements that $E_{\rm rot} > W_{e^{\pm}}$ and that \lhaasosrc's age is $< \tau_{\rm IC}$ (Equation \ref{eqn:ic_timescale}).
We find that a conservative accounting of adiabatic losses requires $P_{0} \lesssim 5$ms.
This conservative upper limit on $P_0$ neglects rotational energy injected after $t=\tau_{\rm SD}$ and the inevitable IC and synchrotron losses would push this upper limit even lower.

The calculations in this section used the fiducial $D_{\gamma} = 2$~kpc distance estimate.
Since $R_{\rm PWN}(\tau_{\rm true}) \propto D$ (Equation \ref{eqn:lhaaso_radius}) and $W_{e^{\pm}} \propto D^{-1}$ (Equation \ref{eq:W_e}), the PWN energy $E_{\rm PWN}(\tau_{\rm sd}) \propto D^{2}$ and our derived initial spin period upper limits scale as $P_{0} \propto D^{-1}$.
$P_0$ also scales with the explosion energy and ejecta mass $P_0 \propto E_{\rm PWN}(\tau_{\rm true})^{-1/2} \propto  R_{\rm PWN}(\tau_{\rm sd})^{1/2} \propto E_{\rm SN}^{3/20}$ and $P_0 \propto E_{\rm PWN}(\tau_{\rm true})^{-1/2} \propto  R_{\rm PWN}(\tau_{\rm sd})^{1/2} \propto M_{\rm ej}^{-1/4}$.
Finally, considering the above dependencies and the statistical uncertainty in $W_{e^{\pm}}$ (Equation \ref{eq:W_e}), we estimate 
\begin{equation}
    P_{0} \lesssim 5_{-2.5}^{+0.9}~\left(\frac{D_{\gamma}}{2~{\rm kpc}}\right)^{-1}~\left(\frac{M_{\rm ej}}{10~M_{\odot}}\right)^{-1/4}~\left(\frac{E_{\rm SN}}{10^{51}~{\rm erg}}\right)^{3/20}{\rm ms}.
\end{equation}

\section{Discussion}
\label{section:discusion}
We consider magnetar \psr \ the most likely candidate to power TeV source \lhaasosrc, and follow-up observations and analysis can confirm this association.
Confirmed magnetar TeV emission has remained elusive in part because magnetars are primarily located in crowded regions of the Galactic plane \citep{ola2014}, where it is difficult to identify the origins of TeV emission from among many possible sources (molecular clouds, supernova remnants, and pulsars).
{\it XMM-Newton} observations of magnetar Swift J1834.9$-$0846 indicate the presence of an X-ray PWN, but confirming its contribution to nearby TeV source HESS J1834$-$097 is complicated by the coincident SNR W41 and a candidate pulsar (XMMU J183435.3$-$084443) that could also be responsible for the TeV $\gamma$-rays \citep{you2012,you2016}. 
Similarly, the magnetar CXOU J171405.7$-$381031 in SNR CTB 37B is suspected to contribute to TeV source HESS J1713$-$381, but the surrounding SNR and nearby molecular clouds are also plausible sources of the observed TeV $\gamma$-rays \citep{hal2010,got2019}.
It is likewise unclear if the $\gamma$-ray source HESS J1808$-$204 is powered by the magnetar SGR 1806$-$20, the nearby massive stellar cluster or a molecular cloud \citep{hess2018}.
This difficulty in identifying magnetar TeV emission is unfortunate because magnetar wind nebulae act as calorimeters, probing otherwise unobservable past magnetar activity.

\subsection{Magnetar B Fields, Initial Spin Periods, and $P-\dot{P}$ Evolution}

The origin of magnetar magnetic fields is not obvious, with various evidence both favoring and disfavoring the competing `fossil field' and dynamo generation hypotheses \citep{spr2008}. 
\cite{dun1992} proposed that fast initial spin periods $P_{0} \sim 1$~ms, could generate magnetar-strength magnetic fields through a dynamo mechanism, and protoneutron star simulations also suggest that a convective dynamo can amplify neutron star dipole magnetic fields to $10^{15}$~G \citep{whi2022}.
However, \cite{vink2006} measured the forward shock radii $r_s$, ages $\tau$, and ISM densities $\rho$ of three SNRs hosting magnetars 1E 1841$-$045, 1E 2259+586, and  SGR 0526$-$66.  % rephrase 
They calculated the explosion energies using the standard Sedov solution
\begin{equation}
R_s^5 = 2.026 \frac{ E_{\rm SN} \tau^2}{\rho},
\end{equation}
and found typical SNR energies $E_{\rm SN} \sim 10^{51}$~erg.
They concluded that if these three magnetars were born with $P_{0} < 5$~ms, then most of the initial rotational energy seems not to have affected their surrounding SNRs.

\begin{figure*}
\centering
\includegraphics[width=0.7\linewidth]{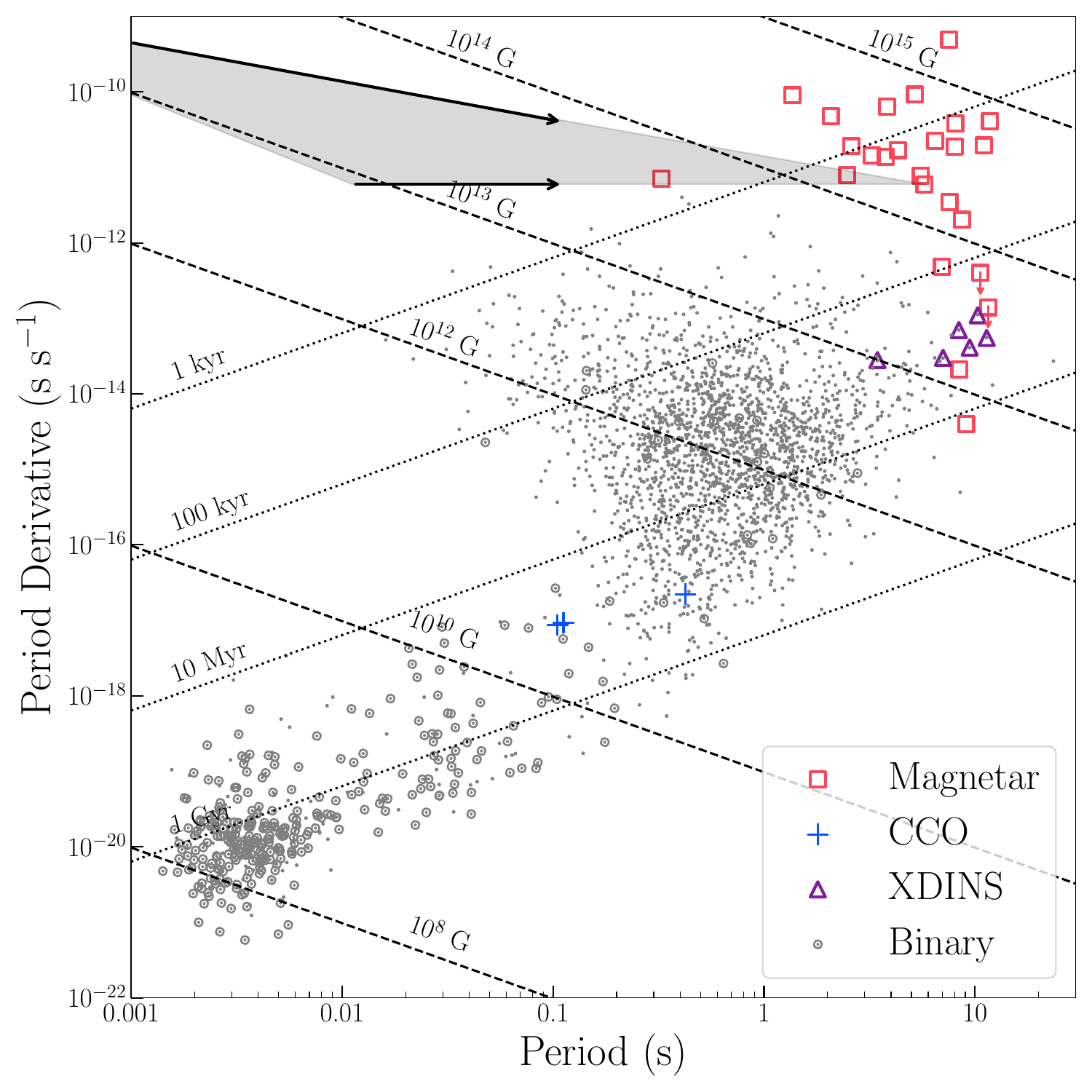}
\caption{Evolution of \psr \ through a $P-\dot{P}$ diagram if it was born with braking index $2.0 \lesssim n \lesssim 2.5$ and spin-down timescale $\tau_{\rm sd} \lesssim 100$~yr, as required by the energy budget constraints calculated in Section \ref{section:energy_budget}.
Data are from the ATNF Pulsar Catalog version 2.1.1 \citep{man2005} and the McGill Online Magnetar Catalog \citep{ola2014}.}
\label{fig:ppot}
\end{figure*}

Figure \ref{fig:ppot} shows \psr's possible trajectories (shaded gray) through the $P$-$\dot{P}$ diagram consistent with the constraints derived in Figure \ref{fig:n_vs_tau}.  
These trajectories correspond to the red curve in the right panel of Figure \ref{fig:n_vs_tau}, with $2.0 \lesssim n \lesssim2.5$ and $\tau_{\rm SD} \lesssim 100$~yr, and \psr's spin-down measured dipole B field increasing over time.
All $P$-$\dot{P}$ trajectories pass close by the positions of PSR J1846$-$0258 in SNR Kes 75 and Swift J1834.9$-$0846, both magnetars with PWNe \citep{got2000,you2012,you2016}.
PSR J1846$-$0258's braking index has been measured to vary from $n=2.65 \pm 0.01$ before an outburst to $n = 2.16 \pm 0.13$ post-outburst \citep{liv2011}, indicating it is \emph{currently} moving roughly parallel to the $P-\dot{P}$ trajectories indicated for \psr.
Evidence of magnetar trajectories through the $P$-$\dot{P}$ diagram with $n <3$ have been derived from proper motion measurements directed away from assumed birth locations in massive star clusters, with $n\sim1.8$ for SGR 1806$-$20 and $n\sim1.2$ for SGR 1900+14 \citep{ten2012}.

\subsection{Directions for Future Work}
JVLA radio follow-up observations can resolve, or place upper limits on, the synchrotron emission from the particles in \lhaasosrc.
This would further constrain the magnetic field strength and morphology of \lhaasosrc, potentially providing further evidence that \psr \ produced a TeV wind nebula by confirming that the magnetic field is low enough to be consistent with \psr's age.
In the hard X-ray band, {\it NuSTAR} data can provide critical information about both the spatially resolved magnetic field strength within \lhaasosrc, and also probe for spatial variations of the maximum $e^{\pm}$ energy throughout \lhaasosrc.

\section{Conclusion}
We have presented evidence that the magnetar \psr \ powers \lhaasosrc.
\psr's position, age, and energetics are consistent with this scenario.
If \psr's rotational energy powers \lhaasosrc, then it must have spun down rapidly ($\tau_{\rm sd} \lesssim 30$~yr) from a fast initial spin period: $P_{0} \lesssim 5~\left(\frac{D_{\gamma}}{2~{\rm kpc}}\right)^{-1}~{\rm ms}$.
This is consistent with theories that magnetar-strength magnetic fields originate from the rapid initial rotation of a protoneutron star.
Follow-up observations and analysis at radio, X-ray, and $\gamma$-ray energies can confirm that \psr \ powers \ \lhaasosrc \ and then further constrain the birth properties of \psr.

\begin{acknowledgements}
We thank Huirong Yan for helpful discussions on cosmic rays.
J.A. would like to thank Jordan Eagle for helpful discussions regarding the Fermi Source Catalog and Fermi LAT data analysis.
The research presented in this paper has used data from the Canadian Galactic Plane Survey, a Canadian project with international partners, supported by the Natural Sciences and Engineering Research Council.
\end{acknowledgements}

\software{{\tt astropy} \citep{ast13, ast18}, {\tt matplotlib} \citep{hun07}, {\tt xspec} \citep{arn96}, {\tt numpy} \citep{har20}, {\tt scipy} \citep{vir20}}

\appendix
\section{Molecular Cloud Scenario Calculations}
\label{appendix_mc}
The angular distance from the center of \lhaasosrc \ to the center of HB9 is $\theta_{\rm HB9} \approx 1.5^{\circ}$.
The \lhaasosrc \ region has volume $V_{\gamma} \propto \theta_{\gamma}^3$.
A spherical shell centered on HB9 with outer radius $\theta_{\rm HB9} + \theta_{\gamma}$ and inner radius $\theta_{\rm HB9} - \theta_{\gamma}$ has a volume $V_{\rm shell} \propto [(\theta_{\rm HB9}+\theta_{\gamma})^3 - (\theta_{\rm HB9}-\theta_{\gamma})^3]$ which implies $V_{\rm shell} / V_\gamma \approx 80$.
The cosmic ray particle energy currently within the spherical shell centered on HB9 is then 
\begin{equation}
W_{\rm shell} = W_{p^{+}} \left(\frac{V_{\rm shell}}{V_\gamma}\right)  \approx 1.1^{+0.2}_{-0.2} \times 10^{50}~\left(\frac{D_{\gamma}}{2~\rm{kpc}}\right)^2~\rm{erg}. 
\end{equation}
The cosmic ray crossing time $\tau_{\rm cross}$ from the inner radius to the outer radius of the  shell is:
\begin{equation}
\tau_{\rm cross} \approx \frac{2 R_{\gamma}}{c} \approx 94 \pm 16~\left(\frac{D_{\gamma}}{2~\rm{kpc}}\right)~{\rm yr},
\end{equation}
which corresponds to power $\dot{E}_{p^{+}}$ injected into cosmic ray protons:
\begin{equation}
\dot{E}_{p} \equiv \frac{W_{\rm shell}}{\tau_{\rm cross}} \approx 3.8_{-1.2}^{+1.2} \times 10^{40}~\left(\frac{D_{\gamma}}{2~\rm{kpc}}\right)~\rm{erg}~\rm{s}^{-1}.
\end{equation}
A $0.7\pm0.4$~kpc distance to HB9 \citep{ran2022} and an age $\tau_{\rm snr}\sim5$~kyr \citep{lea2007} implies that, if HB9 powers \lhaasosrc, then over its lifetime HB9 injected energy $E_{p}\sim \dot{E}_{p} \tau_{\rm snr} \approx 2 \times 10^{51}$~erg into cosmic rays.
This is an order of magnitude larger than HB9's estimated explosion energy $(1.5-3) \times 10^{50}$~erg \citep{lea2007}, and disfavors a hadronic origin for \lhaasosrc.

The LHAASO nondetection of SNR HB9 also disfavors a hadronic origin for \lhaasosrc.
If $\sim70$~TeV $p^{+}$ are accelerated in HB9, then they should be accompanied by similar energy $e^{\pm}$. 
\cite{ara2014} found that Fermi observations of HB9 imply electron exponential cutoff energies $\approx 500$~GeV in the IC scenario and proton exponential cutoff energies $\approx 80$~GeV in the hadronic scenario.
These exponential cutoffs in the particle spectra within HB9 in both the leptonic and hadronic scenarios are consistent with the LHAASO nondetection of HB9, and both cutoff energies are much lower than the $E_{\rm cut} \approx 70$~TeV inferred for \lhaasosrc \ (Table \ref{table:SED}).
This disfavors a hadronic origin for \lhaasosrc.

\section{SNR Scenario Calculations}
\label{appendix_snr}

The leptonic particle spectrum inferred from IC modeling of the \lhaasosrc \ SED (Table \ref{table:SED}) would emit a higher synchrotron X-ray flux unless
\begin{equation}
\label{B_snr_e}
B_{\rm snr} \lesssim 10~{\rm \mu G}  \ {\rm (leptonic \ case)},
\end{equation}
assuming $D_\gamma = 2~{\rm kpc}$ and closer distances requiring smaller $B_{\rm snr}$.
If \lhaasosrc \ is a previously unknown SNR accelerating protons up to at least $\sim70$~TeV (cf. Table \ref{table:SED}), then:
\begin{equation}
\label{B_snr_p}
B_{\rm snr} \lesssim 35~{\rm \mu G} \ {\rm (hadronic \ case)},
\end{equation}
which we have calculated by assuming a cosmic ray proton to electron ratio $K_{\rm ep} = 0.01$.
We will discuss both leptonic and hadronic scenarios, and consider an SNR currently in the free expansion or Sedov phase in Section \ref{section:possible_power_sources:SNR_free_expansion}, and an SNR currently in the radiative phase in Section \ref{section:possible_power_sources:SNR_radiative_phase}.

\subsection{SNR Scenario: Free Expansion or Sedov Phase}
\label{section:possible_power_sources:SNR_free_expansion}
If \lhaasosrc \ is a previously unknown SNR accelerating electrons up to a maximum energy $ E_{\rm e,max} \sim40$~TeV, then either $E_{\rm e,max}$ is limited the SNR age, synchrotron losses or particle escape
If \lhaasosrc \ is a previously unknown SNR accelerating protons up to  $ E_{\rm p,max} \sim70$~TeV,  then either $E_{\rm p,max}$ is limited by either the SNR age or particle escape.
\cite{rey1998} calculated the theoretical maximum particle energy in each of these scenarios: 
\begin{equation}
E_{\rm max} (\rm TeV) = \begin{cases}
                \eta^{-1}~B_{10}~V_{16}^2~t_{250}, & \text{age \ limited} \\
                \eta^{-1}~B_{10}^{-1/2}~V_{16}, & \text{loss \ limited} \\
                B_{10}~\lambda, & \text{escape \ limited} 
                \end{cases} 
\end{equation}
In the above equations we have defined $t_{250} \equiv \left(t_{\rm snr} / 250~{\rm yr}\right)$, $V_{16} \equiv \left(V_{\rm s} / 16000~{\rm km~s^{-1}}\right)$, $B_{10} \equiv \left(B_{\rm snr} / 10~{\rm{\mu G}}\right)$ and the maximum MHD wavelength $\lambda \equiv (\lambda_{\rm max} / 2.5 \times 10^{17}~{\rm cm})$.
The $B_{\rm snr}$ dependence of the maximum energies, and our upper limit on $B_{\rm snr}$ (Equations \ref{B_snr_e} and \ref{B_snr_p}) , allow us to constrain the minimum required shock velocity in each case.
A Sedov phase SNR must have a shock speed $V_{\mathrm{s}} \gtrsim 1.2 \times 10^4$~km~s$^{-1}$, while a freely expanding SNR requires an even larger  $V_{\mathrm{s}}$.

$V_s$ corresponds to proton temperature (see e.g. \cite{vin2012}):
\begin{equation}
kT_{\mathrm{p}} \approx  \frac{3}{16}  m_{p} V_{\mathrm{s}}^{2} \approx 196~\mathrm{keV}~\left(\frac{V_s}{10^{4}~\mathrm{km~s^{-1}}}\right)^{2}.
\label{eqn:snr_temperature_p}
\end{equation}
The Coulomb equilibration timescale $\tau_{ep}$ for the electrons to reach thermal equilibrium with the protons is (\cite{vin2012,ito1984,zel1966}):
\begin{equation}
\tau_{ep} \approx 10~\left(\frac{n_{p}}{1 \ \rm{cm}^{-3}}\right)^{-1}~\left(\frac{\overline{kT}}{1~\rm{keV}}\right)^{3/2}~\left(\frac{\rm{ln}~\Lambda}{30.9}\right)^{-1}~\rm{kyr},
\end{equation}
where  $n_{\rm p}$ is  the proton density, the mean temperature $\overline{kT} \approx 0.6~kT_{\mathrm{p}}$, and $\rm{ln}~\Lambda$ is the Coulomb logarithm:
\begin{equation}
\rm{ln}~\Lambda = 30.9 - \rm{ln}\left[n_e^{1/2} \left(\frac{kT_e}{1~\rm{keV}}\right)^{-1} \right].
\end{equation}
For all $V_s \gtrsim 1.2 \times 10^4$~km~s$^{-1}$, $\tau_{ep}$ is much greater than a typical SNR age, so we cannot calculate the thermal X-ray emission from this hypothetical SNR by setting the electron temperature equal to $kT_{\mathrm{SNR}}$.
\citep{gha2007,gha2013}
We instead multiply $kT_{\mathrm{p}}$ by the electron to proton mass ratio to get a hard lower bound on the electron temperature: 
\begin{equation}
kT_{\mathrm{e}} \gtrsim \left(\frac{m_e}{m_p}\right) kT_{\mathrm{p}}  \approx 107~\mathrm{eV}  \left(\frac{V_s}{1.2 \times10^{4}~\mathrm{km~s^{-1}}}\right)^{2}
\label{eqn:snr_temperature_e}
\end{equation}
Observations of SNRs with shock velocities $V_s \gtrsim 400$~km~s$^{-1}$ indicate $kT_{\mathrm{e}} \gtrsim 0.3$~keV  \citep{gha2007,gha2013}, so Equation \ref{eqn:snr_temperature_e} is probably unrealistically low, but we will adopt it here for the most conservative X-ray flux calculation.
We use the {\tt apec} model available in {\tt Xspec} to calculate the 0.1$-$2.5~keV flux from a thermal plasma with temperature $kT_{\mathrm{e}} =107~\mathrm{eV}$ \citep{arn96}.
The model normalization is: 
\begin{equation}
N = \frac{10^{-14}}{4 \pi R_\gamma^2} \int n_{\rm e} n_{\rm H}   \rm{d}V
\end{equation}
and we have set $n_{\rm e} = 1.2~n_{\rm H}$ (assuming 
cosmic abundances) and $\int \rm{d}V = \frac{4}{3} \pi R_\gamma^3$.
We find that $n_e \lesssim 0.02~{\rm cm^{-3}}$ is required to match the unabsorbed ROSAT X-ray flux upper limit $F_x$ (Equation \ref{eq:rosat_flux}).
A Sedov phase SNR origin for \lhaasosrc \ is therefore ruled out by the archival ROSAT upper limits on its X-ray emission, as long as the SNR electron density $n_e \gtrsim 0.02~{\rm cm^{-3}}$ (and a free expansion phase SNR would require an even lower $n_e$).

An SNR in the Sedov phase has radius $R_{\rm Sedov}$ \citep{dra2011}:
\begin{equation}
 \label{eq:sedov_radius}
R_{\rm Sedov} = 5.0~{\rm pc}~E_{51}^{1/5} n_{0}^{-1/5} t_{3}^{2/5} 
\end{equation}
and $V_{\rm Sedov}$:
\begin{equation}
 \label{eq:sedov_velocity}
V_{\rm Sedov} = 1950~{\rm km~s^{-1}}~E_{51}^{1/5} n_{0}^{-1/5} t_{3}^{-3/5}.
\end{equation}
Requiring the Sedov shock velocity $V_{\rm Sedov} > 12000~\mathrm{km~s^{-1}}$ at the beginning of the Sedov phase and also requiring $n_{0} \lesssim 0.02~{\rm cm^{-3}}$ implies that $E_{51} > 5$.
This low density environment and high explosion energy would be atypical.

\subsection{SNR Scenario: Radiative Phase SNR}
\label{section:possible_power_sources:SNR_radiative_phase}
 An SNR transitions from the Sedov phase to the radiative phase at time $t_{\rm rad}$ \citep{dra2011,der2009,blo1998}:
 \begin{eqnarray}
 \label{eq:radiative_time}
t_{\rm rad} &=& 49.3~{\rm kyr}~E_{51}^{0.22}~n_{0}^{-0.55} \\
&\approx&      13.9~{\rm kyr}~E_{51}^{0.22}~\left(\frac{n_{0}}{10~{\rm cm^{-3}}}\right)^{-0.55},
\end{eqnarray}
and the transition occurs at radius $R_{\rm rad}$:
\begin{eqnarray}
R_{\rm rad} &=& 23.7~{\rm pc}~E_{51}^{0.29}~n_{0}^{-0.42} \\
 &\approx& 9~{\rm pc}~E_{51}^{0.29}~\left(\frac{n_{0}}{10~{\rm cm^{-3}}}\right)^{-0.42},
\end{eqnarray}
when the SNR has a forward shock velocity $V_{\rm rad}$:
\begin{eqnarray}
V_{\rm rad} &=& 188~{\rm km~s^{-1}}~\left(E_{51}~n_{0}^2\right)^{0.07} \\
 &\approx& 260~{\rm km~s^{-1}}~E_{51}^{0.07}~\left(\frac{n_{0}}{10~{\rm cm^{-3}}}\right)^{0.14}.
\end{eqnarray}
If \lhaasosrc \ is powered by an SNR that is currently in the radiative phase, 
 the material behind the forward shock \emph{might} have cooled enough that its thermal X-ray flux is below the ROSAT upper limit (Equation \ref{eq:rosat_flux}).

If \lhaasosrc \ is associated with a radiative SNR, then the SNR must have had sufficiently fast shocks to accelerate electrons above $\approx40$~TeV in the past and the SNR must have also reached the radiative phase before the electrons could cool below $\approx40$~TeV today.
The magnetic field inside a radiative SNR is expected to at least exceed the typical ISM value, i.e. $B_{\rm snr} \gtrsim 2~\mu$G, which implies the combined IC and synchrotron cooling timescale is $\lesssim 15$~kyr.
Setting $t_{\rm rad} < 15$~kyr in Equation \ref{eq:radiative_time} requires:
\begin{eqnarray}
E_{51}^{0.22}~\left(\frac{n_{0}}{10~{\rm cm^{-3}}}\right)^{-0.55} < 1.08.
\end{eqnarray}
Since this leptonic radiative SNR scenario must invoke fast shocks, at least at the beginning of the Sedov phase, Equation \ref{eq:sedov_velocity} requires: 
\begin{eqnarray}
E_{51}^{1/5}~\left(\frac{n_{0}}{10~{\rm cm^{-3}}}\right)^{-1/5} \gtrsim 2.
\end{eqnarray}
For a typical explosion energy $E_{51} = 1$ the two equations above are equivalent to the contradictory requirements that $n_0 \gtrsim 9~{\rm cm^{-3}}$ and  $n_0 \lesssim 0.3~{\rm cm^{-3}}$, and these two contradictory  $n_0$ lower/upper limits differ by at least a factor of 7 for any $10^{-3} < E_{51} < 10$.
This inconsistency disfavors a leptonic, radiative SNR origin for \lhaasosrc.

Finally, we consider the possibility that \lhaasosrc \ is hadronic and associated with a currently radiative SNR.
The diffusion coefficient for $E \sim 70$~TeV protons is:
\begin{eqnarray}
\mathbb{D} &=& \mathbb{D}_0 \left(\frac{E}{1~\rm{GeV}}\right)^{\delta} \\
&\approx& 3 \times 10^{29}~{\rm cm}^2~{\rm s}^{-1},
\end{eqnarray}
where we have set $\delta = 1/3$ (corresponding to a Kolmogorov spectrum of magnetic turbulence), 
and $\mathbb{D}_0 = 4 \times 10^{28}$~cm$^2$~s$^{-1}$, a typical value in the Milky Way \citep{str2007}.
The corresponding diffusion length $l$ in a time $t$ is: 
\begin{eqnarray}
l &=& \sqrt{4\mathbb{D}t} \\
&\approx& 580~\left(\frac{t}{25~{\rm kyr}}\right)^{0.5}~{\rm pc}.  
\end{eqnarray}
This is a least $\approx 40$ times larger than the observed \lhaasosrc \ radius $R_\gamma$ for any distance $D_\gamma \leq 2$~kpc. 
We conclude that if \lhaasosrc \ is hadronic and powered by an SNR that is now radiative then the local diffusion coefficient $\mathbb{D}_0 \lesssim  1.8  \times~10^{25}$~cm$^2$~s$^{-1}$, at least three orders of magnitude lower than the typical value in the Milky Way.

In comparison, HAWC observations of the Geminga and Monogem pulsar halos indicate that the local  diffusion coefficient for the 100~TeV~$e^{\pm}$ in these regions are $4.5 \pm 1.2 \times~10^{27}$~cm$^2$~s$^{-1}$, corresponding to $\mathbb{D}_0 \approx  2.2  \times~10^{26}$~cm$^2$~s$^{-1}$ for $\delta = 1/3$ \citep{abe2017}.
The Geminga and Monogem observations may indicate a $e^{\pm}$ diffusion coefficient around these PWNe that is lower than the typical value in the Milky Way, but these are observations of $e^{\pm}$, not protons. 
It is not clear if an even lower diffusion coefficient for protons propagating away from an old radiative SNR could be consistent with other observations of cosmic ray protons in the Milky Way, such as their inferred galactic escape timescale and isotropy.


\begin{thebibliography}{}

\bibitem[Abeysekara et al.(2017)]{abe2017} Abeysekara, A.~U., Albert, A., Alfaro, R., et al.\ 2017, Science, 358, 911. doi:10.1126/science.aan4880

\bibitem[Ackermann et al.(2013)]{ack2013} Ackermann, M., Ajello, M., Allafort, A., et al.\ 2013, Science, 339, 807. doi:10.1126/science.1231160

\bibitem[Aharonian(2013)]{aha2013} Aharonian, F.~A.\ 2013, Astroparticle Physics, 43, 71. doi:10.1016/j.astropartphys.2012.08.007

\bibitem[Araya(2014)]{ara2014} Araya, M.\ 2014, \mnras, 444, 860. doi:10.1093/mnras/stu1484

\bibitem[Archibald et al.(2016)]{arc2016} Archibald, R.~F., Gotthelf, E.~V., Ferdman, R.~D., et al.\ 2016, \apjl, 819, L16. doi:10.3847/2041-8205/819/1/L16

\bibitem[Arnaud(1996)]{arn96} Arnaud, K.~A.\ 1996, Astronomical Data Analysis Software and Systems V, 101, 17. 

\bibitem[Astropy Collaboration et al.(2013)]{ast13} Astropy Collaboration, Robitaille, T.~P., Tollerud, E.~J., et al.\ 2013, \aap, 558, A33. doi:10.1051/0004-6361/201322068

\bibitem[Astropy Collaboration et al.(2018)]{ast18} Astropy Collaboration, Price-Whelan, A.~M., Sip{\H{o}}cz, B.~M., et al.\ 2018, \aj, 156, 123. doi:10.3847/1538-3881/aabc4f

\bibitem[Ballet et al.(2023)]{bal2023} Ballet, J., Bruel, P., Burnett, T.~H., et al.\ 2023, arXiv:2307.12546. doi:10.48550/arXiv.2307.12546

\bibitem[Beloborodov \& Li(2016)]{bel2016} Beloborodov, A.~M. \& Li, X.\ 2016, \apj, 833, 261. doi:10.3847/1538-4357/833/2/261

\bibitem[Blanchard et al.(2020)]{bla2020} Blanchard, P.~K., Berger, E., Nicholl, M., et al.\ 2020, \apj, 897, 114. doi:10.3847/1538-4357/ab9638

\bibitem[Blondin et al.(1998)]{blo1998} Blondin, J.~M., Wright, E.~B., Borkowski, K.~J., et al.\ 1998, \apj, 500, 342. doi:10.1086/305708

\bibitem[Bolatto et al.(2013)]{bol2013} Bolatto, A.~D., Wolfire, M., \& Leroy, A.~K.\ 2013, \araa, 51, 207. doi:10.1146/annurev-astro-082812-140944

\bibitem[Camero et al.(2014)]{cam2014} Camero, A., Papitto, A., Rea, N., et al.\ 2014, \mnras, 438, 3291. doi:10.1093/mnras/stt2432

\bibitem[Cao et al.(2024)]{cao2024} Cao, Z., Aharonian, F., An, Q., et al.\ 2024, \apjs, 271, 25. doi:10.3847/1538-4365/acfd29

\bibitem[Chevalier(1977)]{che1977} Chevalier, R.~A.\ 1977, Supernovae, 66, 53. doi:10.1007/978-94-010-1229-4\_5

\bibitem[Chrimes et al.(2025)]{chr25} Chrimes, A.~A., Levan, A.~J., Lyman, J.~D., et al.\ 2025, \aap, 696, A127. doi:10.1051/0004-6361/202453479

\bibitem[Dame et al.(2001)]{dam2001} Dame, T.~M., Hartmann, D., \& Thaddeus, P.\ 2001, \apj, 547, 792. doi:10.1086/318388

\bibitem[Dame \& Thaddeus(2022)]{dam2022} Dame, T.~M. \& Thaddeus, P.\ 2022, \apjs, 262, 5. doi:10.3847/1538-4365/ac7e53

\bibitem[Dere et al.(2009)]{der2009} Dere, K.~P., Landi, E., Young, P.~R., et al.\ 2009, \aap, 498, 915. doi:10.1051/0004-6361/200911712

\bibitem[Draine(2011)]{dra2011} Draine, B.~T.\ 2011, Physics of the Interstellar and Intergalactic Medium by Bruce T. Draine. Published by Princeton University Press, 2011. ISBN: 978-0-691-12214-4

\bibitem[Duncan \& Thompson(1992)]{dun1992} Duncan, R.~C. \& Thompson, C.\ 1992, \apjl, 392, L9. doi:10.1086/186413

\bibitem[Gelfand et al.(2009)]{gel2009} Gelfand, J.~D., Slane, P.~O., \& Zhang, W.\ 2009, \apj, 703, 2051. doi:10.1088/0004-637X/703/2/2051

\bibitem[Ghavamian et al.(2007)]{gha2007} Ghavamian, P., Laming, J.~M., \& Rakowski, C.~E.\ 2007, \apjl, 654, L69. doi:10.1086/510740

\bibitem[Ghavamian et al.(2013)]{gha2013} Ghavamian, P., Schwartz, S.~J., Mitchell, J., et al.\ 2013, \ssr, 178, 633. doi:10.1007/s11214-013-9999-0

\bibitem[Gotthelf et al.(2000)]{got2000} Gotthelf, E.~V., Vasisht, G., Boylan-Kolchin, M., et al.\ 2000, \apjl, 542, L37. doi:10.1086/312923

\bibitem[Gotthelf et al.(2019)]{got2019} Gotthelf, E.~V., Halpern, J.~P., Mori, K., et al.\ 2019, \apj, 882, 173. doi:10.3847/1538-4357/ab378c

\bibitem[Green(2019)]{gre2019} Green, D.~A.\ 2019, Journal of Astrophysics and Astronomy, 40, 36. doi:10.1007/s12036-019-9601-6

\bibitem[Green(2025)]{gre2025} Green, D.~A.\ 2025, Journal of Astrophysics and Astronomy, 46, 14. doi:10.1007/s12036-025-10055-9

\bibitem[Halpern \& Gotthelf(2010)]{hal2010} Halpern, J.~P. \& Gotthelf, E.~V.\ 2010, \apj, 725, 1384. doi:10.1088/0004-637X/725/1/1384

\bibitem[Harris et al.(2020)]{har20} Harris, C.~R., Millman, K.~J., van der Walt, S.~J., et al.\ 2020, \nat, 585, 357. doi:10.1038/s41586-020-2649-2

\bibitem[H.~E.~S.~S. Collaboration et al.(2018)]{hess2018} H.~E.~S.~S. Collaboration, Abdalla, H., Abramowski, A., et al.\ 2018, \aap, 612, A11. doi:10.1051/0004-6361/201628695

\bibitem[HI4PI Collaboration et al.(2016)]{HI2016} HI4PI Collaboration, Ben Bekhti, N., Flöer, L., et al.\ 2016, \aap, 594, A116. doi:10.1051/0004-6361/201629178

\bibitem[Hobbs et al.(2004)]{hob2004} Hobbs, G., Lyne, A.~G., Kramer, M., et al.\ 2004, \mnras, 353, 1311. doi:10.1111/j.1365-2966.2004.08157.x

\bibitem[Hobbs et al.(2005)]{hob2005} Hobbs, G., Lorimer, D.~R., Lyne, A.~G., et al.\ 2005, \mnras, 360, 3, 974. doi:10.1111/j.1365-2966.2005.09087.x

\bibitem[Hunter(2007)]{hun07} Hunter, J.~D.\ 2007, Computing in Science and Engineering, 9, 90. doi:10.1109/MCSE.2007.55

\bibitem[Itoh(1984)]{ito1984} Itoh, H.\ 1984, \apj, 285, 601. doi:10.1086/162535

\bibitem[Jing et al.(2023)]{jin2023} Jing, W.~C., Han, J.~L., Hong, T., et al.\ 2023, \mnras, 523, 4949. doi:10.1093/mnras/stad1782

\bibitem[Kaspi \& Beloborodov(2017)]{kas2017} Kaspi, V.~M. \& Beloborodov, A.~M.\ 2017, \araa, 55, 261. doi:10.1146/annurev-astro-081915-023329

\bibitem[Leahy \& Tian(2007)]{lea2007} Leahy, D.~A. \& Tian, W.~W.\ 2007, \aap, 461, 1013. doi:10.1051/0004-6361:20065895

\bibitem[Livingstone et al.(2011)]{liv2011} Livingstone, M.~A., Ng, C.-Y., Kaspi, V.~M., et al.\ 2011, \apj, 730, 66. doi:10.1088/0004-637X/730/2/66

\bibitem[Olausen \& Kaspi(2014)]{ola2014} Olausen, S.~A. \& Kaspi, V.~M.\ 2014, \apjs, 212, 6. doi:10.1088/0067-0049/212/1/6

\bibitem[Ranasinghe \& Leahy(2022)]{ran2022} Ranasinghe, S. \& Leahy, D.\ 2022, \apj, 940, 63. doi:10.3847/1538-4357/ac940a

\bibitem[Reynolds(1998)]{rey1998} Reynolds, S.~P.\ 1998, \apj, 493, 375. doi:10.1086/305103

\bibitem[Reynolds \& Chevalier(1984)]{rey1984} Reynolds, S.~P. \& Chevalier, R.~A.\ 1984, \apj, 278, 630. doi:10.1086/161831

\bibitem[Rybicki \& Lightman(1979)]{ryb1979} Rybicki, G.~B. \& Lightman, A.~P.\ 1979, A Wiley-Interscience Publication, New York: Wiley, 1979

\bibitem[Manchester et al.(2005)]{man2005} Manchester, R.~N., Hobbs, G.~B., Teoh, A., et al.\ 2005, \aj, 129, 1993. doi:10.1086/428488

\bibitem[Mong \& Ng(2018)]{mon2018} Mong, Y.-L. \& Ng, C.-Y.\ 2018, \apj, 852, 86. doi:10.3847/1538-4357/aa9e90

\bibitem[Spruit(2008)]{spr2008} Spruit, H.~C.\ 2008, 40 Years of Pulsars: Millisecond Pulsars, Magnetars and More, 983, 391. doi:10.1063/1.2900262

\bibitem[Strong et al.(2007)]{str2007} Strong, A.~W., Moskalenko, I.~V., \& Ptuskin, V.~S.\ 2007, Annual Review of Nuclear and Particle Science, 57, 285. doi:10.1146/annurev.nucl.57.090506.123011

\bibitem[Tan et al.(2020)]{tan2020} Tan, C.~M., Bassa, C.~G., Cooper, S., et al.\ 2020, \mnras, 492, 5878. doi:10.1093/mnras/staa113

\bibitem[Tendulkar et al.(2012)]{ten2012} Tendulkar, S.~P., Cameron, P.~B., \& Kulkarni, S.~R.\ 2012, \apj, 761, 76. doi:10.1088/0004-637X/761/1/76

\bibitem[Thaddeus(1977)]{tha1977} Thaddeus, P.\ 1977, Star Formation, 75, 37

\bibitem[Vink \& Kuiper(2006)]{vink2006} Vink, J. \& Kuiper, L.\ 2006, \mnras, 370, L14. doi:10.1111/j.1745-3933.2006.00178.x

\bibitem[Vink(2012)]{vin2012} Vink, J.\ 2012, \aapr, 20, 49. doi:10.1007/s00159-011-0049-1
\bibitem[Virtanen et al.(2020)]{vir20} Virtanen, P., Gommers, R., Oliphant, T.~E., et al.\ 2020, Nature Methods, 17, 261. doi:10.1038/s41592-019-0686-2

\bibitem[White et al.(2022)]{whi2022} White, C.~J., Burrows, A., Coleman, M.~S.~B., et al.\ 2022, \apj, 926, 111. doi:10.3847/1538-4357/ac4507

\bibitem[Xu et al.(2006)]{xu2006} Xu, Y., Reid, M.~J., Zheng, X.~W., et al.\ 2006, Science, 311, 54. doi:10.1126/science.1120914

\bibitem[Younes et al.(2012)]{you2012} Younes, G., Kouveliotou, C., Kargaltsev, O., et al.\ 2012, \apj, 757, 39. doi:10.1088/0004-637X/757/1/39

\bibitem[Younes et al.(2016)]{you2016} Younes, G., Kouveliotou, C., Kargaltsev, O., et al.\ 2016, \apj, 824, 138. doi:10.3847/0004-637X/824/2/138

\bibitem[Zabalza(2015)]{zab2015} Zabalza, V.\ 2015, 34th International Cosmic Ray Conference (ICRC2015), 34, 922. doi:10.22323/1.236.0922

\bibitem[Zeldovich \& Raizer(1966)]{zel1966} Zeldovich, Y.~B. \& Raizer, Y.~P.\ 1966, New York: Academic Press, 1966, edited by Hayes, W.D.; Probstein, Ronald F.

\end{thebibliography}
\end{document}